\documentclass[10pt,preprint2,a4paper]{aastex}
\usepackage[left=2.1cm,top=2cm,right=0.8cm,nohead,nofoot]{geometry}
\usepackage{graphics,epsf}
\usepackage{amsmath}                
\usepackage{amsfonts}               
\usepackage{amssymb}                
\usepackage{epsfig}                 

\def \kev{\rm{keV}}

\def \cm{~\rm{cm}}
\def \s{~\rm{s}}
\def \km{~\rm{km}}

\def \AU{~\rm{AU}}

\def \yr{~\rm{yr}}

\def \kpc{~\rm{kpc}}

\def \days{~\rm{day}}

\def \keV{~\rm{keV}}
\def \astrobj#1{#1}

\begin{document}

\title{THE ORIENTATION OF THE ETA CARINAE BINARY SYSTEM}

\author{Amit Kashi\altaffilmark{1} and Noam Soker\altaffilmark{1}}

\altaffiltext{1}{Department of Physics, Technion$-$Israel Institute of
Technology, Haifa 32000 Israel; kashia@physics.technion.ac.il;
soker@physics.technion.ac.il.}

\setlength{\columnsep}{1cm}
\small

\begin{abstract}
We examine a variety of observations that shed light on the orientation
of the semi-major axis of the $\eta$ Carinae massive binary system.
Under several assumptions we study the following observations: The
Doppler shifts of some He~I P-Cygni lines that is attributed to the
secondary's wind, of one Fe~II line that is attributed to the primary's
wind, and of the Paschen emission lines which are attributed to the
shocked primary's wind, are computed in our model and compared with
observations. We compute the hydrogen column density toward the binary
system in our model, and find a good agreement with that deduced from
X-ray observations. We calculate the ionization of surrounding gas
blobs by the radiation of the hotter secondary star, and compare with
observations of a highly excited [Ar III] narrow line. We find that all
of these support an orientation where for most of the time the
secondary$-$the hotter less massive star$-$is behind the primary star.
The secondary comes closer to the observer only for a short time near
periastron passage, in its highly eccentric ($e\simeq 0.9$) orbit.
Further supporting arguments are also listed, followed by discussion of
some open and complicated issues.
\end{abstract}

\keywords{ (stars:) binaries: general$-$stars: mass loss$-$stars:
winds, outflows$-$stars: individual ($\eta$ Car)}

\section{INTRODUCTION}
\label{sec:intro}

The $P=5.54 \yr$ ($P =2022.7 \pm 1.3~$d; Damineli et al. 2008a)
periodicity of the massive binary system \astrobj{$\eta$ Car} is
observed in the radio (Duncan \& White 2003), IR (Whitelock et al.
2004), visible (e.g., van Genderen et al. 2006), X-ray (Corcoran 2005),
and in many emission and absorption lines (e.g., Damineli et al. 2008a,
b). According to most models the periodicity, e.g., of the
spectroscopic event and of the X-ray minimum, follows the 5.54~years
periodic change in the orbital separation in this highly eccentric,
$e \simeq 0.9$, binary system (e.g., Hillier et al. 2006). The
spectroscopic event is defined by the fading, or even disappearance, of
high-ionization emission lines (e.g., Damineli 1996; Damineli et al.
1998, 2000, 2008a,b; Zanella et al. 1984). The rapid changes in
the continuum, lines, and in the X-ray properties (e.g., Martin et al.
2006,a,b; Davidson et al. 2005; Nielsen et al. 2007; van Genderen et
al. 2006; Damineli et al. 2008b; Corcoran 2005) are assumed to occur
near periastron passages, although less rapid variations occur along
the entire orbit.

It is generally agreed that the orbital plane lies in the equatorial
plane of the bipolar structure$-$the Homunculus (Davidson et al. 2001).
The inclination angle (the angle between a line perpendicular to the
orbital plane and the line of sight) is $i \simeq 45 ^\circ$, with
$i=41^\circ-43^\circ$ being a popular value (Davidson et al. 2001;
Smith 2002). However, there is a disagreement about the orientation of the
semimajor axis in the orbital plane$-$the periastron longitude. We will
use the commonly used periastron longitude angle $\omega$:
$\omega=0^\circ$ for a case when the secondary is toward the observer
at an orbital angle of $90^\circ$ after periastron,
$\omega=90^\circ$ for a case when the secondary is toward the observer
at periastron, $\omega=180^\circ$ for a case when the secondary is
toward the observer at an orbital angle $90^\circ$ before periastron,
and $\omega=270^\circ$ for a case when the secondary is toward the
observer at apastron, and so on.

While some groups argue that the secondary (less massive) star is away
from us during periastron passages, $\omega=270^\circ$ (e.g., Nielsen
et al. 2007; Damineli et al. 2008b), others argue that the secondary is
toward us during periastron passages, $\omega=90^\circ$
(Falceta-Gon\c{c}alves et al. 2005; Abraham et al. 2005; Kashi \& Soker
2007b [hereafter KS07], who use the angle $\gamma=90^\circ-\omega$).
Other semimajor axis orientations have also been proposed (Davidson
1997; Smith et al. 2004; Dorland 2007; Henley et al. 2008; Okazaki et
al. 2008a).
Abraham \& Falceta-Gon\c{c}alves (2007) have obtained the orientation angle in the range
$\omega=60- 90^\circ$, independent of the orbital inclination.
They did not assume that the binary and the Homunculus orbital plane coincide.
This contradicts the binary interacting model we support.
In addition, Falceta-Gon\c{c}alves et al. (2005) suggested a model for the X-ray emission
with the usage of a problematic expression for the X-ray emission (see
Akashi et al. 2006). We therefore will not use the arguments of these
two papers to support our claim for $\omega \simeq 90^\circ$.

In the present paper we consider all observations that can shed light
on the periastron longitude. As will be shown, all of them support a value
of $\omega \simeq 90^\circ$. Namely, the secondary is behind the
primary most of the time, and passes in front of the primary for a short
time near periastron passage. In section \ref{sec:doppler} we discuss
the Doppler shifts of several lines, in section \ref{sec:N_H} we
discuss the hydrogen column density as deduced from X-ray observations,
and section \ref{sec:narrow} contains a discussion of the variation in
the intensity of high excitation narrow lines. Our discussion and
predictions are in section \ref{sec:diss}.

\begin{figure}[!t]
\resizebox{0.49\textwidth}{!}{\includegraphics{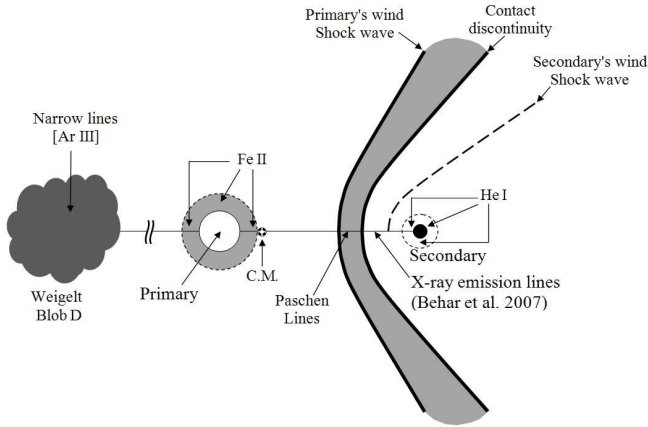}}
\caption{\footnotesize A schematic map of the $\eta$ Car system at apastron.
The origins of the different lines are marked on the map.
According to our suggested model ($\omega=90^\circ$),
the observer is on the left side of the system.}
\label{fig:map}
\end{figure}

Our assumptions concerning the location where some lines are produced is
different from those assumed by other research groups.
We therefore present in Figure \ref{fig:map} a map of the origins of the different
lines according to our assumptions.
This figure should be consulted when reading the sections to follow.

\section{THE DOPPLER SHIFTS AS ARISING FROM ORBITAL MOTION}
\label{sec:doppler}

\subsection{The binary system}
\label{subsec:binary}

In applying the orbital motion we will use two sets of binary parameters.
In both the orbital period is taken to be 2024 days
(in a recent paper the orbital period was found to be 2021-2024 day;
Damineli et al. 2008a), and the eccentricity is very high $e \sim 0.9$.
The terminal winds' velocities are taken to be $v_1=500 \km \s^{-1}$ and $v_2=3000 \km \s^{-1}$, where
subscripts $1$ and $2$ stand for the primary and secondary,
respectively. The first binary model is based on the commonly used
binary parameters as composed from many papers (e.g.,
Ishibashi et al. 1999; Damineli et al. 2000; Corcoran et al. 2001,
2004b; Hillier et al. 2001; Pittard \& Corcoran 2002; Smith et al.
2004; Verner et al. 2005; Damineli et al. 2008a, b; Dorland 2007). This
is termed by us (Kashi \& Soker 2008b, hereafter KS08) the
`small-masses model'. The assumed stellar masses are $M_1=120 M_\odot$,
$M_2=30 M_\odot$. The calculated semimajor axis is $a=16.64 \AU$,
and the orbital separation at periastron is $r=1.66-1.16 \AU$ for
$e=0.9-0.93$, respectively.

The masses of the small-masses model are fixed more or less by the
assumption that each star is at its Eddington luminosity limit.
However, fitting to evolution of main sequence stars gives much
higher masses. This model is termed the `big-masses model'. Following
KS08, we take masses that come from evolutionary considerations (see
also Damineli et al. 2000), and are in the ranges $M_1 \simeq 150-180
M_\odot$, as the primary has lost several tens of solar masses, and $M_2
\simeq 60-70 M_\odot$.

In KS07 we fixed the stellar masses and varied the inclination angle
$i$ and the periastron longitude $\omega$ (using the angle
$\gamma=90^\circ-\omega$). In the present paper we fix $i=45^\circ$,
$e=0.9$ or $e=0.93$, and $\omega=90^\circ$ (secondary toward us at periastron), and
use two sets of masses: $(M_1,M_2)=(120,30)M_\odot$ and
$(M_1,M_2)=(160,60)M_\odot$. We will not play at all with the
parameters, hence the fitting will not be perfect. Small variations of
the parameters $i$, $\omega$, $M_1$, $M_2$, and $e$ can easily give
better fits. But until we know better the values of these parameters by
other means, it is pointless to play with these five parameters. What
we do stress is that the orbital motion can explain the observed
Doppler shifts.

\subsection{Observations Supporting Orbital Motion Interpretation}
\label{subsec:support}

There are several arguments supporting the interpretation of the
periodic variations in the Doppler shifts as arising from the orbital
motion. However, the interpretation is not straightforward. Damineli et
al. (1997, 2000; see also Davidson 1997) attributed the Doppler shift
of the hydrogen emission Paschen lines to the orbital motion of the
primary star. Using this interpretation they found the eccentricity to
be $e=0.75$, much lower that the now commonly assumed value of $e\simeq
0.9$; Davidson (1997) found $e \simeq 0.7-0.8$. In addition, this
amplitude of the Doppler shift if originating in one of the stars is
not compatible with other Doppler shifts as we discuss in this paper.
Alternatively, we will attribute the Paschen line emission to the
shocked primary's wind near the stagnation point of the colliding
winds.

By attributing some of the He~I P Cygni lines to the secondary$-$the
less massive star$-$we were able to explain the Doppler periodicity
with the commonly used binary parameters (KS07). We consider this as a
strong support to the orbital motion interpretation, and devote section
\ref{subsec:dop} to study this in more detail. There we also show that
the Fe~II~$\lambda6455$ might be coming from the primary's wind. If our
interpretation is correct, $\eta$ Car is a double-lined spectroscopic
binary. We note that our view that some of the He~I lines originate in
the secondary's wind is in dispute, as most researchers in the field
attribute the He~I P Cygni lines to other sources, either the primary
or the winds collision region (Falceta-Gon\c{c}alves et al. 2007;
Humphreys et al. 2008). Our reply to some of the criticisms of
Humphreys et al. (2008) are in section 5 of KS07, where we also discuss
why attributing the He~I lines to the primary's wind is problematic.
In particular, if the change in the Doppler shift of the absorbed
part was due to the change in the distance of the wind from the primary
surface where the lines are formed, hence a change in the wind
velocity, we would expect the lines emission width to change
accordingly, and its center to stay at the stellar primary velocity.
This is contrary to observations. Take the He~I~$\lambda7067$
line for example, as reported by Nielsen et al. (2007); the
center of the emission part follows the minimum in the line profile,
and its width does not change much. This is what is expected if the
entire source of the line changes its velocity, as in the orbital
interpretation of the Doppler shifts.

In our model the emission of the $\lambda7065$, $\lambda5876$, and
$\lambda4471$ He~I lines are formed in the secondary's wind. The
secondary luminosity of $\sim 20 \%$ of the primary luminosity seems to
be enough for forming these lines (see KS07). We attribute the
absorption in these lines$-$the P-Cygni profile$-$to the secondary
wind. The absorption is from the emission in this line. Indeed, the
minimum of the absorption in the $\lambda7065$ and $\lambda5876$ lines
is clearly above the continuum of the $\eta$ Car spectrum (Hillier et
al. 2001). In the case of the He~I~$\lambda4471$ line it is less clear.
However, we note that the entire region from 4400 to 4500$\rm{\AA}$ is
full of lines (e.g. Pereira et al. 2008), that form a region above the
real continuum. Some of these lines are formed in the secondary star.
It is therefore possible that the continuum is actually 10 percent
lower than the value used by Hillier et al. (2001), and that the
He~I~$\lambda4471$ line has its absorption above the continuum as well.
We conclude that even though the helium lines show strong absorption,
it is still above the continuum, and therefore it is possible that the
lines are emitted and absorbed in the secondary's wind, as we suggest
here.

Another support to the orbital motion interpretation might come from
the silicon and sulfur X-ray lines. If the lines are taken to be formed
where the fast secondary's wind is shocked, then their Doppler shift
variations with orbital phase can be explained by using the same
orbital parameters as used to explain the Doppler shifts of the He~I
lines (Behar et al. 2008). Henely et al. (2008) studied the possibility
that the Doppler shifts of these X-ray lines result mainly from the
outflow velocity of the shocked secondary's wind. As can be seen from
their Figure 18 this model fails. Although the outflow velocity of the
shocked secondary is likely to contribute to the Doppler shift, the
main variations in the center of the lines must be explained by other
means, e.g., the orbital motion as we suggest.

The Fe~II~$\lambda 6455$ line also shows variations in the
Doppler shift toward the central source (Damineli et al. 2008b), but
not in the polar-direction reflected light (Stahl et al. 2005; this
line will be studied in section \ref{subsec:dop}).

\subsection{Fitting the Doppler Shift with Orbital Motion}
\label{subsec:dop}

\begin{figure}[!t]
\resizebox{0.49\textwidth}{!}{\includegraphics{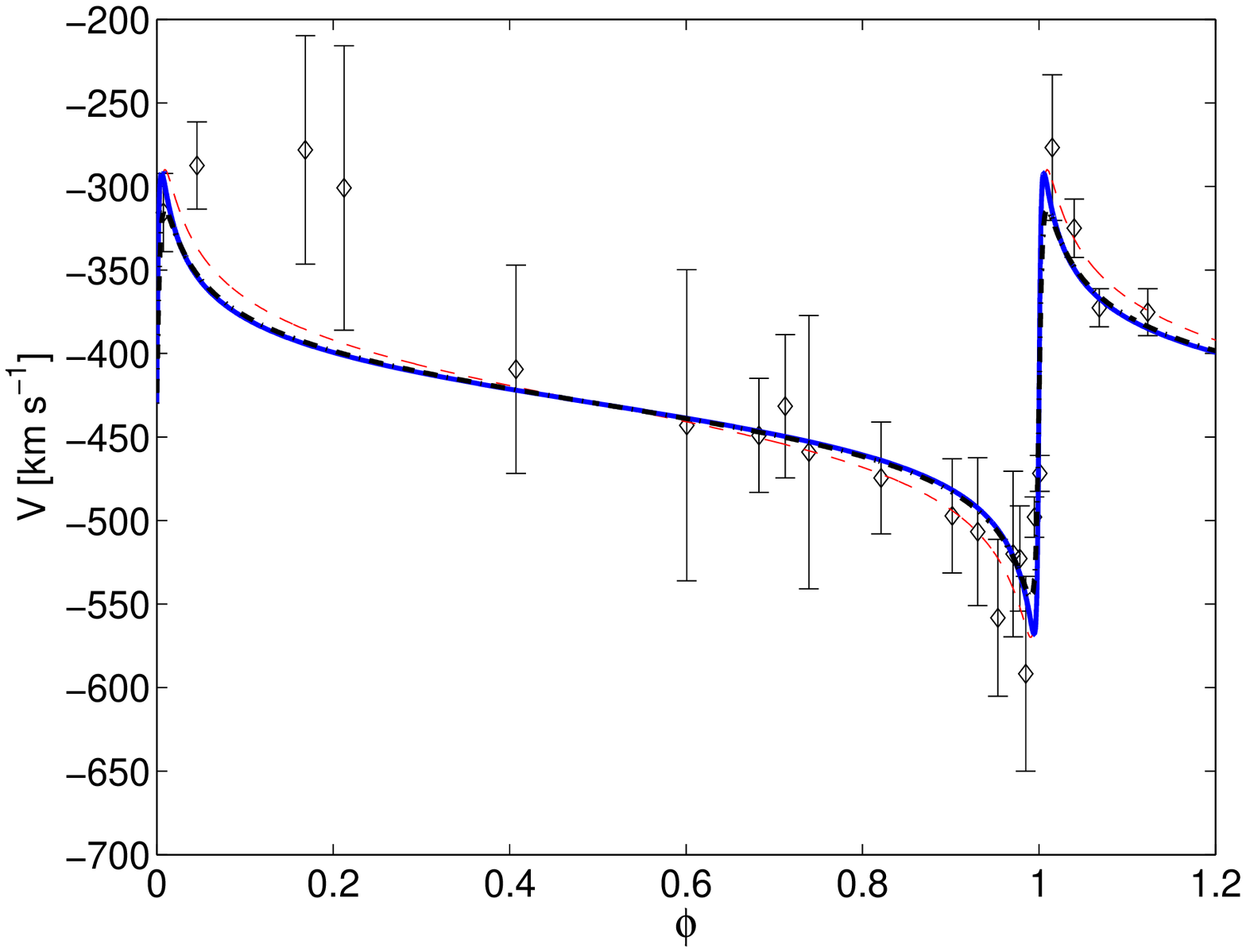}}
\resizebox{0.49\textwidth}{!}{\includegraphics{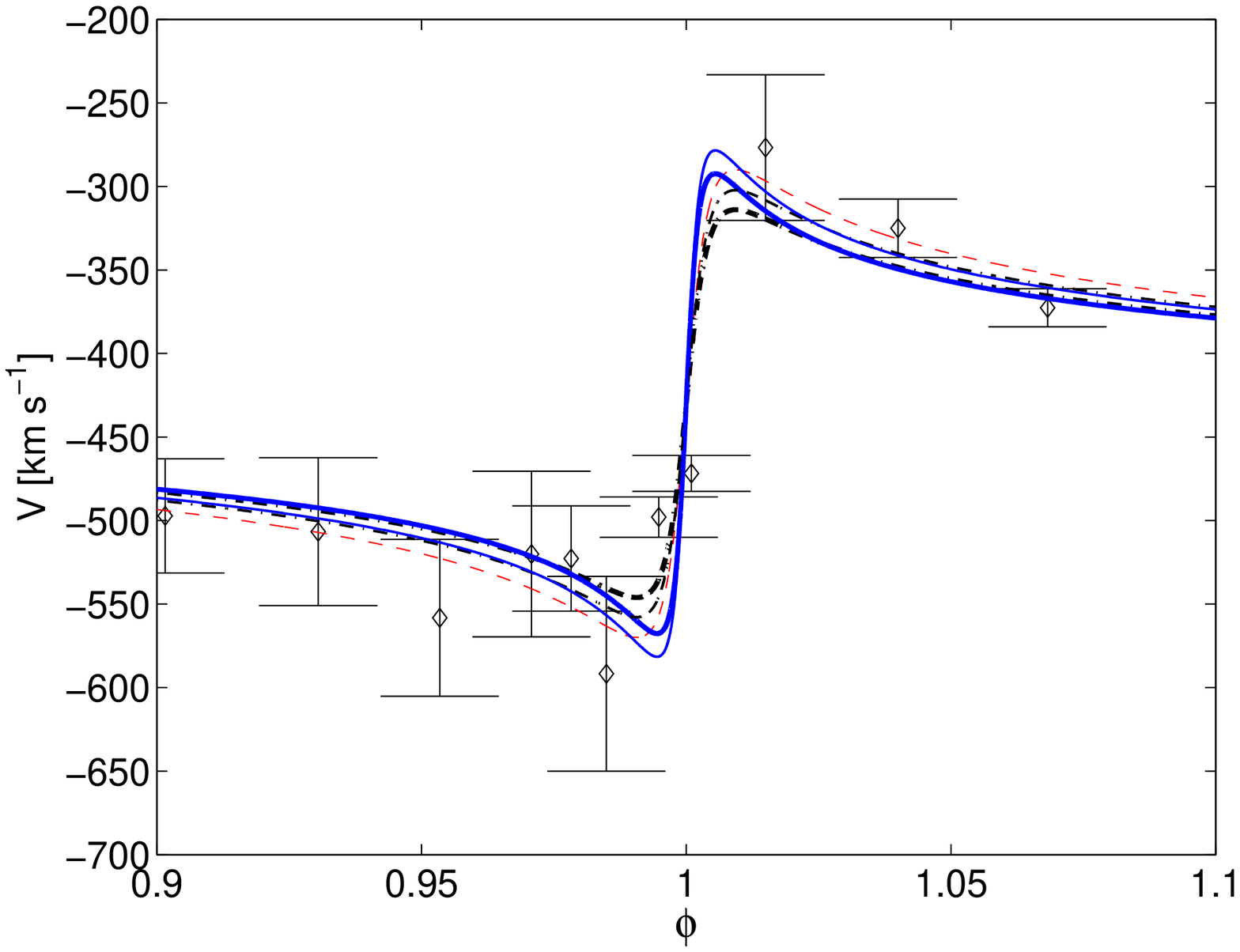}} 
\caption{\footnotesize The Doppler shift of the He~I lines. In our
model the source of the He~I lines is the acceleration zone of the
secondary's wind. The inclination angle is $i=45^\circ$,
$\omega=90^\circ$ (i.e., secondary toward us at periastron) for both
panels. Diamonds and error bars: He~I lines data from Nielsen et al.
(2007). Red thin line: The mathematical fit made by Nielsen et al.
(2007). Blue line presents results for eccentricity of $e=0.90$, and
the black dot-dashed line for $e=0.93$. First panel: the small-masses
model, $(M1,M2)=(120,30)M_\odot$. Second panel: zoom close to
periastron passage. The big-masses model, $(M1,M2)=(160,60)M_\odot$ is
also presented (thin lines) in the second panel.}
\label{fig:HeI}
\end{figure}

We start by presenting the main results of KS07, and extending them to
the big-masses model. The assumption is that lines with a difference
between minimum and maximum of more than $200 \km \s^{-1}$ arise from
the secondary star, or the acceleration zone of its wind.
We attribute the He~I lines formation to the acceleration zone of the
secondary's wind. Our model is based on a simple Doppler shift. The
parameters are: (a) Inclination, (b) Masses of two stars (c)
Eccentricity (d) Orbital period. All these are taken from other works
in the literature--we only changed the inclination and eccentricity by
several percents. (e) The secondary's wind region where the lines are
formed. This is the only new parameter in our model. By fitting the
line formation region to $v=-430 \rm{km/s}$ we got the best fit
(KS07). Noticeable are some of the broad He~I lines. These lines are
formed in the dense part of the wind, namely, where the velocity is far
from its terminal speed. Typically the velocity there is in the range
$400-500 \km \s^{-1}$, while the secondary's wind terminal velocity is
$\sim 3000 \km \s^{-1}$. In the first panel of Figure \ref{fig:HeI} we
present two cases of the small-masses model; more examples for this
model are in KS07. As the data are noisy, a fit should be used. We did
not add the $-8 \km \s^{-1}$ system velocity of $\eta$ Car as it is
negligible relative to other uncertainties at this stage. Our
calculations are compared with the data from Nielsen et al. (2007).
Nielsen et al. (2007) do not present a physical model, but rather a
pure mathematical fitting to the data points. Being so close to Nielsen
et al. (2007) model is considered favorable to our model and our
suggested orientation.

The orbital motion alone cannot explain in full the Doppler variations,
and the other effects related to the intrinsic properties of the winds
must be included in a full study. The ionizing radiation of the
secondary's wind influences the primary's wind acceleration zone (e.g.,
Nielsen et al. 2007; Damineli et al. 2008b), and the radiation of the
colliding winds might influence the acceleration zone of the
secondary's wind (Soker \& Behar 2006). The influence of these types of
radiation increases with decreasing orbital separation, and is not
equal at similar phases before and after phase zero (periastron). These
are the main effects related to the variations in lines intensity.
However, compared to the Doppler shift these are second order effects.
This is supported by the observation
of intensity variations toward the polar direction, but without Doppler
variations (see section \ref{subsec:support} here). For that, our fit, in
particular near periastron, cannot be perfect, as these effects are not
included in the present study.
Considering this, and that we did not try to play with the parameters too much,
we consider the results presented in Figure \ref{fig:HeI} as a strong support
to the orbital motion interpretation of the Doppler shifts, and for a values of
$\omega=90^\circ$.
We note that Nielsen et al. (2007) and Damineli et al. (2008b) attributed the
P Cygni lines to the primary's wind, and their Doppler shift variations to the
influence of the secondary's ionizing radiation.

\begin{figure}[!t]
\resizebox{0.49\textwidth}{!}{\includegraphics{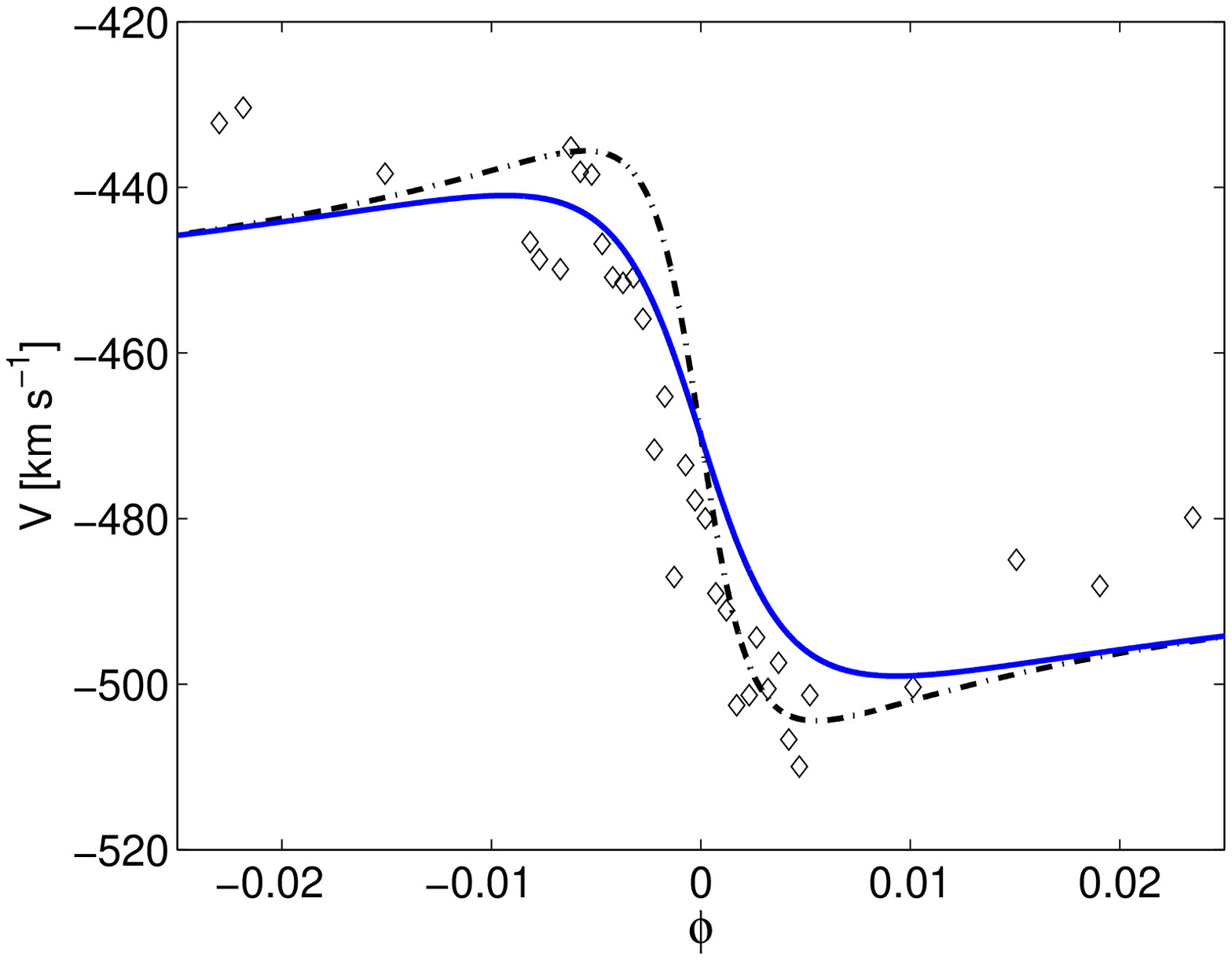}}
\resizebox{0.49\textwidth}{!}{\includegraphics{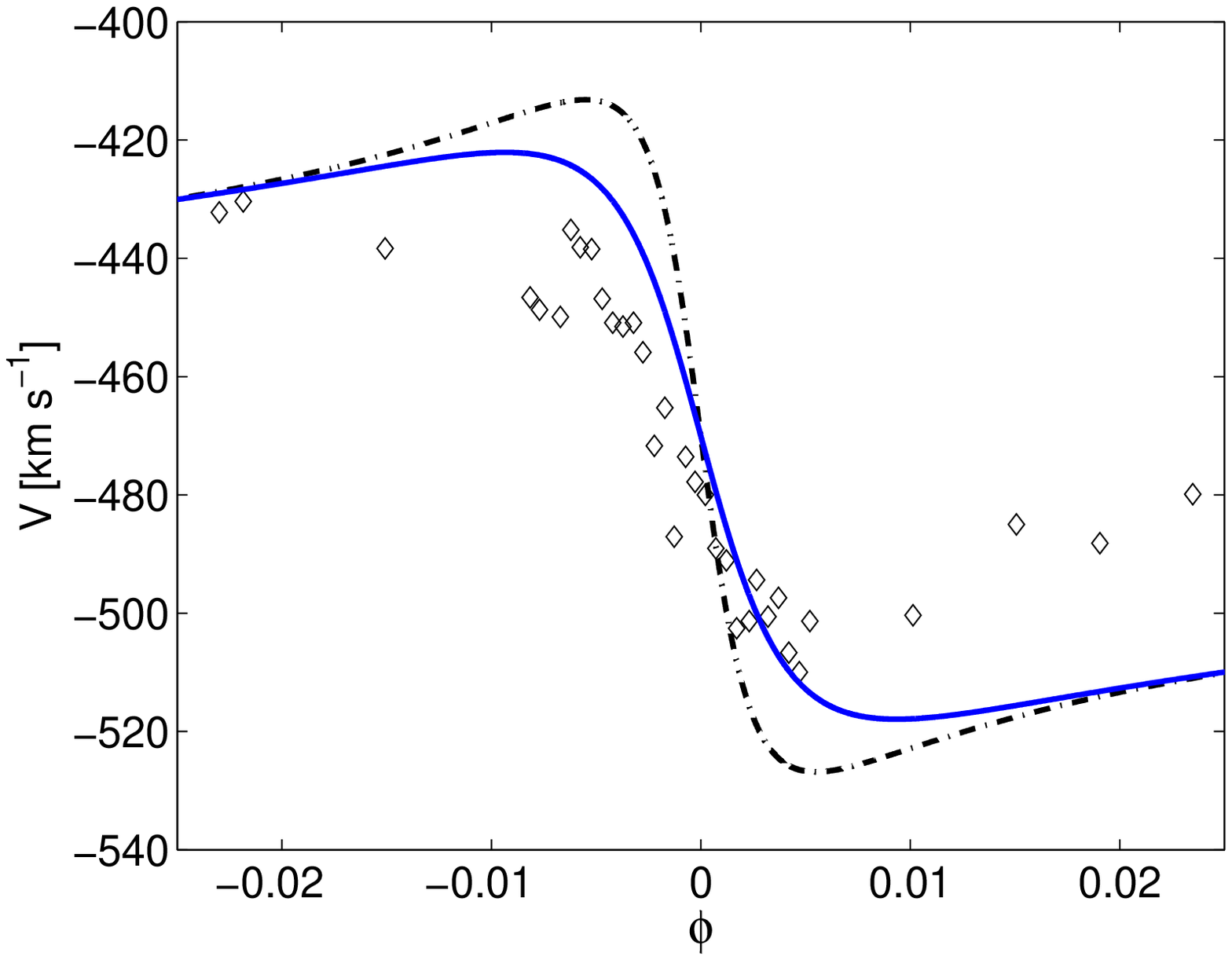}}
\caption{\footnotesize The Doppler shift of the Fe~II~$\lambda6455$ line.
In our model the source of the
Fe~II~$\lambda6455$ line is the primary's wind. The inclination
angle is $i=45^\circ$ for both panels. Diamonds:
Fe~II~$\lambda6455$ line data from Damineli et al. (2007b). Blue
line: $e=0.90$. Black dot-dashed line: $e=0.93$. First panel-
$(M1,M2)=(120,30)M_\odot$. Second panel-
$(M1,M2)=(160,60)M_\odot$.} \label{fig:FeII}
\end{figure}

Some lines show much lower periodic velocity variations, and are likely to
be formed in other regions of the binary system and its winds.
The first candidate is the primary and its wind.
We attribute the Fe~II~$\lambda6455$ line to the primary's wind.
Using the same binary parameters as for the lines arising from the secondary,
and in particular the same orientation with $\omega=90^\circ$,
namely the primary is on the far side at periastron and toward us at apastron,
we calculated the Doppler shift of this line.
Our results compared with the observations of Damineli et al. (2007b)
near periastron passage are presented in Figure \ref{fig:FeII}. As
discussed above it is unlikely to perfectly fit the shifts near
periastron. However, our interpretation gives correctly the amplitude
of the velocity variation, and suggest that this line can arise in the
primary's wind.

The colliding wind region can also be a source of lines showing orbital
Doppler shifts.
Behar et al. (2007) have already considered the Doppler shift of X-ray emission lines
from the shocked secondary's wind region. This region is closer to the
center of mass than the secondary is, hence the amplitude of the
Doppler shift is lower. Still closer to the center of mass is the
shocked primary's wind along the line joining the two stars. This region
is only a few$\times 0.01 r_{2s}$ closer to the center of mass than the
stagnation point is, where $r_{2s}$ is the distance of the secondary
star from the stagnation point (Kashi \& Soker 2007a).
We therefore take this region to be at a distance $0.02 r_{2s}$ closer to
the center of mass (closer to the primary) than the stagnation point is.
We will try to attribute the Doppler shift of the hydrogen Paschen emission lines to this region.
Damineli et al. (1997, 2000) and Davidson (1997) already noticed that the Doppler
shifts of these lines might follow an eccentric orbit behavior, but they attributed its
source to the primary star.

\begin{figure}[!t]
\resizebox{0.49\textwidth}{!}{\includegraphics{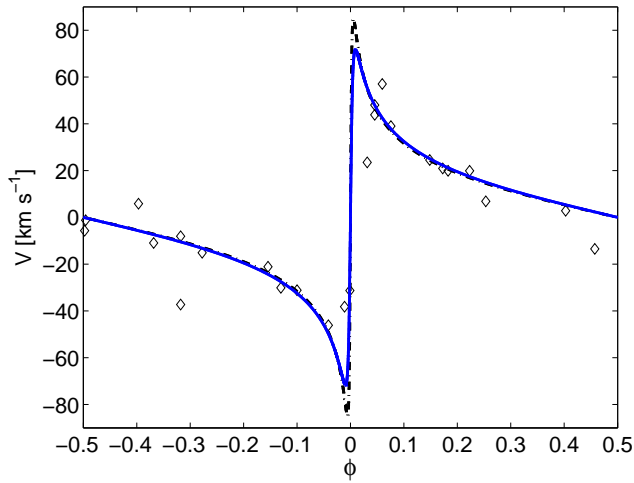}}
\caption{\footnotesize The Doppler shift of the Paschen lines. In our
model the source of the Paschen lines is the shocked primary's wind
which is $0.02 r_{2s}$ closer to the center of mass (closer to the
primary) than the stagnation point is, where $r_{2s}$ is the distance
of the secondary star from the stagnation point. The inclination angle
is $i=45^\circ$ for both panels. Diamonds: Paschen lines data from
Damineli et al. (2000; unfortunately error bars were not available).
Blue line: our model with $e=0.90$. Black dot-dashed line: our model with $e=0.93$
(Differences between the lines can be noticed only close to periastron).}
\label{fig:paschen}
\end{figure}

In Figure \ref{fig:paschen} we plot the observations of Damineli et al.
(2000), as well as the expected Doppler shift of lines emitted from a
region near the stagnation point in our binary model.
Our calculation fits very nicely the Doppler shift.
Only just before and after phase zero our calculated Doppler shift overestimate
the observed value. In any case, because of the ionization by the two stars
(see discussion above), and the expected collapse of the winds interaction region,
we would be surprise if the simple model fit the observations very close to phase zero.
Near periastron passages the colliding winds region might collapse onto
the secondary (Soker 2005; KS08). Namely, this region ceases to
exist in its simple form, and other regions become the dominant source of the Paschen lines.
Although the fit is not perfect, it is good enough considering we did not change any
parameter to fit the Doppler shifts of the hydrogen lines, but rather used the same
parameters as we have done in explaining the He~I and Fe~II lines.
The figure shows our results for the small-masses model $(M1,M2)=(120,30)M_\odot$.
Our results for the big-masses model, $(M1,M2)=(160,60)M_\odot$, which are not presented are
very similar to these of the small-masses model.

Let us elaborate more on the nature of the wind interaction.
The location of the stagnation point is determined by the equilibrium
between the ram pressures of the two winds. The orbital motion
influences the ram pressure, mainly that of the slower primary's wind.
Therefore, when the two stars approach each other the stagnation point
moves closer to the secondary star. Namely, its distance from the
center of mass increases. The opposite behavior occurs as the distance
between the two stars increases after periastron passages. This effect
can be seen in Figure 2 of Akashi et al. (2006). The effect is
significant near periastron. We neglect this asymmetrical behavior
before and after periastron passage for the following reasons.
(1) Even if the Paschen lines formed indeed in the shocked primary's wind, we
don't know where exactly they are formed, and how this region changes
its location as the shocked primary's wind density increases with
decreasing distance. (2) Near periastron many other effects are likely
to occur. The radiations from the two stars influence the colliding
winds; gravity, e.g., tidal interaction, can also influence the
colliding wind process.
(3) An accretion process might occur for $\sim 10$~weeks (KS08).
(4) Due to the orbital motion of the stars the apex of the colliding
winds$-$the stagnation point$-$will not be exactly on the line joining the
centers of the two stars (Soker 2005). The angular offset is too small to be considered in
our analytical study, and we must neglect it.
For all these effects that add to the uncertainties near periastron
passages, and are not considered by us, there is no justification to add the asymmetry in
the stagnation point location to the present first order calculations.
We neglect all these effects in the present study, and in addition don't
really try to fit the parameters to the Paschen lines.

\section{THE COLUMN DENSITY}
\label{sec:N_H}

In this section we discuss the hydrogen column density as deduced
from X-ray observations.
Hamaguchi et al. (2007) calculated the hydrogen column density toward the X-ray
emitting gas in the center of $\eta$ Car. They differentiated between the
column density toward gas having a temperature of $kT>5 \keV$ and a gas
of $kT \sim 1 \keV$. We will study only the emission from the $kT>5 \keV$ gas, because
we know this region comes from the shocked secondary's wind, before it suffers any
adiabatic cooling. Therefore, we can safely estimate its location to be close to the
stagnation region, but somewhat closer to the secondary.

The source and location of the $kT \sim 1 \keV$ gas, on the other hand, is
less secure. It might come from a fast polar outflow from the primary
star (Smith et al. 2004 argue for a fast polar wind from the
primary star). Alternatively, it might come from an oblique shock of
the secondary's wind away from the secondary. Yet, another possibility is
a postshocked secondary's wind that had cooled adiabatically as it streamed
away from the stagnation point region. The very important thing we do learn from
Hamaguchi et al. (2007) analysis of the X-ray emission by the $kT \sim 1 \keV$
gas is that the contribution of gas around the interaction region
to the column density is $N_{H-e} \la 5 \times 10^{22} \cm^{-2}$.
This implies that any column density of $N_H \ga 5 \times 10^{22} \cm^{-2}$ must
come from material close to the binary system, at most $\sim 100 \AU$.

Indeed, the column density toward the $kT>5 \keV$ gas at phase 0.47 is
$N_H=17\times 10^{22} \cm^{-2}$. This is hard to explain in a model where
the secondary is toward as during this phase, i.e., $\omega \sim 270^\circ$.
This is because then our line of sight toward the shocked secondary's wind will
go through the undisturbed secondary's wind, namely, through the opening of the
conical shock formed by the colliding wind. Instead, we suggest that the secondary star
reside on the far side during apastron, and the column density includes the undisturbed
primary's wind as well as the shocked primary's wind.

In principle, if the line of sight goes through the shocked
primary's wind,{{{i.e. $\omega=270^\circ$,}}} then absorption can be very high.
However, if this was the case near apastron, then at a later phase the line of sight
would have gone through the primary's wind before the wind was shocked.
In this case the calculations of Akashi et al. (2006) in their equation
(6) show that the column density becomes $N_H=17\times 10^{22}
\cm^{-2}$ only at an orbital separation of $a=10\AU$, at phase -0.06
(or 0.94). Therefore, {{{orientation with $\omega=270^\circ$}}} will not explain the
value of $N_H=17\times10^{22} \cm^{-2}$ at phase 0.92 as observed by Hamaguchi et al. (2007).
A very specific orientation and wind properties would be required to account
for a large $N_H$ both at phase 0.5 and at phase 0.92 in the case the
secondary is toward us at apastron ($\omega=270^\circ$).
Up to date, there is no specific model showing this.
On the contrary, a new paper by Okazaki et al. (2008b) shows that near apastron the
expected opening angle of the conical shell is larger than the angle of the line of
sight from the equatorial plane ($90^{\circ} - i$), and therefore we
observe the X-ray emitting region through the low density secondary
wind, as we assume.

\begin{figure}[!t]
\resizebox{0.49\textwidth}{!}{\includegraphics{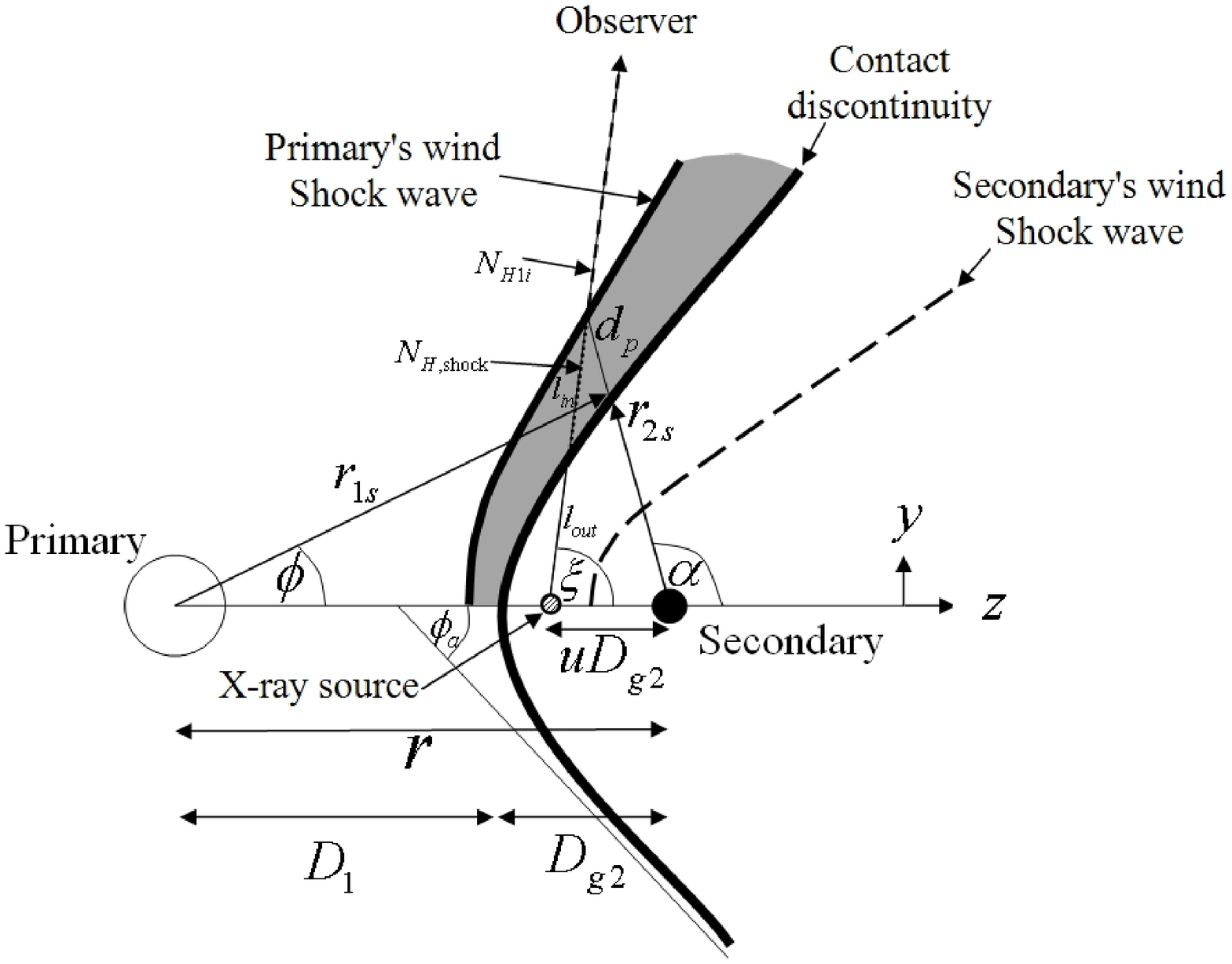}}
\caption{\footnotesize Schematic drawing of the collision region of the
two stellar winds and the definition of several quantities used in the
paper. The point marked `X-ray source' in the post shocked secondary
wind is where we take the $kT>5 \keV$ gas to reside. From there we
calculate the value of $N_H$.} 
\label{fig:NHgeo}
\end{figure}

To calculate the column density for our preferred orientation of  $\omega=90^\circ$
we use the geometry drawn schematically in Figure \ref{fig:NHgeo} for the colliding winds
(for more detail see Kashi \& Soker 2007a, 2008a).
The contact discontinuity shape is approximated by an hyperbola at distance $D_{g2}\simeq0.3 r$
from the secondary at the stagnation point of the colliding winds,
where the two winds momenta balance each other along the symmetry axis,
and with an asymptotic angle of $\phi_a=60^\circ$.

The density of the post-shocked primary's wind introduces large
uncertainties, because it is strongly influenced by the magnetic field
(Kashi \& Soker 2007a).
We shall therefore limit the ram pressure compression factor to the
smaller magnetic pressure compression factor ($f_m$). The thickness of
the post-shocked primary's wind shell (from the contact discontinuity to
the primary's wind shockwave) is calculated accordingly. The hydrogen
number density is
\begin{equation}
n_H = \frac{0.43f_m \dot M_1}{4 \pi r_{1s}^2 v_1 \mu m_H },
\label{eq:nH}
\end{equation}
where $v_{\rm wind1}$ is the primary's wind velocity.
The compression factor together with a few more
assumptions allow us to calculate the velocity of the post shock primary's wind out from the
shock region, and the width of the conical shock. From the width we can calculate the values
of $d_p$ and $l_{in}$ defined in Figure \ref{fig:NHgeo}.

The source of the hard X-ray is the post-shocked secondary's wind
(Corcoran 2005; Akashi et al. 2006), taken here at the point
marked on Figure \ref{fig:NHgeo} by `X-ray source'; of course, the
X-ray emitting region is more extended, and our treatment is crude.
This point is located at a distance of $(1-u)D_{g2}$ from the
stagnation point, or a distance of $uD_{g2}$ from the secondary. We
assume $u = 0.7$. As evident from the figure, the column density has
two main components: The post-shocked primary's wind component,
$N_{H,\rm{shock}}$ (the conical shell), and the undisturbed,
free-expanding, primary's wind component ($N_{Hi1}$). We calculate the
contribution of each component to the total column density
($N_{H,\rm{tot}}$) as a function of orbital angle $\theta$

As before we take the inclination angle to be $i=45^\circ$, and assume that
the secondary is away from us during a apastron passage ($\omega=90^\circ$).
This geometry explicitly determines the direction from which the system is observed
(i.e. line of sight) at each orbital phase.
For every orbital angle $\theta$ we calculate the relevant
direction angle $\xi$ to the observer. Considering the orientation of
the conical shell at that orbital angle, we calculate the thickness of
the conical shell in that direction and integrate $n_H$
over the width to find the column density of the first component:
\begin{equation}
N_{H,\rm{shock}} = \int_{l_{\rm{out}}}^l n_H\,dl \label{eq:NHshock}
\end{equation}
where $l=l_{\rm{out}}+l_{\rm{in}}$.
The second component contributing to the column density is calculated from the
point on the line of sight where the shock terminates to infinity (contribution decreases fast
with distance).
\begin{equation}
N_{Hi1} = \int_l^\infty n_H\,dl \label{eq:NHi1}.
\end{equation}
To calculate the total column density we added a constant values of
$4 \times 10^{22} \cm^{-2}$ for the material residing in the outer regions, e.g.,
in the Homunculus and in the ISM.

\begin{figure}[!t]
\resizebox{0.49\textwidth}{!}{\includegraphics{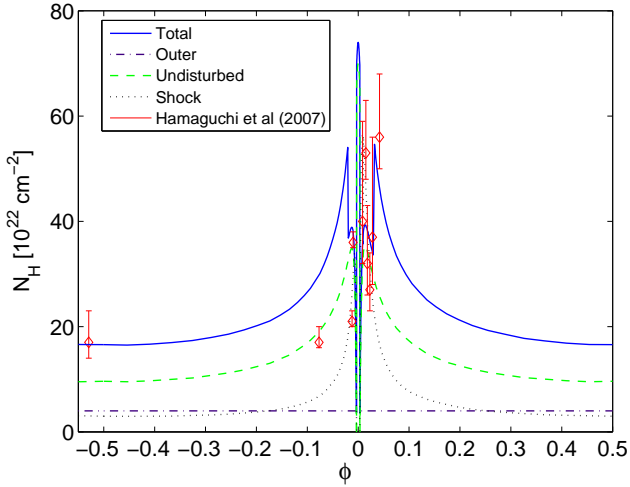}}
\caption{\footnotesize The hydrogen column density obtained from our
model. The dashed-green line represents the undisturbed
(free-expanding) primary's wind component ($N_{Hi1}$); The dotted-black
line represents the post-shocked primary's wind component
($N_{H,\rm{shock}}$); The dot-dashed purple line represents the constant column
density from the Homunculus;
The solid-blue line is the sum of all three components.
{}From phase $-0.02$ to $0.03$ the total column density does not include the
contribution of the conical shell, because our simple model breaks up in that
region (see text). This is the reason for the change in behavior of the
total column density near periastron.
The $N_H$ toward the $kT >5 \keV$ gas from Table 5 of Hamaguchi et
al. (2007) is plotted as red diamonds.
} \label{fig:NH}
\end{figure}

The three column density components and the total column density are
plotted in Figure \ref{fig:NH}. The $N_H[>5 \keV]$ component from
Hamaguchi et al. (2007) is also plotted.

At phase $-0.015$ the relative orbital velocity of the two stars is
large, $\sim 200 \km \s^{-1}$, and the winding of the conical shell
(Nielsen et al. 2007; Okazaki et al. 2008a), as well as the collapse of
the colliding wind region (Kashi \& Soker 2008b), cannot be neglected
until after the recovery at phase $\sim 0.03$ (60 days after periastron
passage). For that, from phase $-0.02$ to $0.03$ the total column
density does not include a contribution from the conical shell, and the
undisturbed primary's wind is assumed to extended to the secondary,
instead of toward the stagnation point. This is the reason for the
change in behavior of the total column density near periastron. In any
case, close to periastron our treatment is very crude, but still gives
the observed values. An exact treatment requires 3D numerical
hydrodynamical calculation, which is beyond the scope of this paper.
The simulations of Okazaki et al. (2008b) are isothermal, and therefore
are not accurate enough to describe the shocked secondary's wind. We
emphasize that the purpose of this calculation is only to point out
that the region of X-ray emission is somewhere near the secondary, and
absorption can be crudely estimated by accounting for these
three absorption components.

We do not take into account the absorption through the unshocked
secondary's wind close to periastron. There are two reasons for that:
The first is that there is no conical shell near periastron. Instead,
the secondary's wind covers a short distance in the radial direction
away from the primary
 because the conical shell (winds-interaction region) is wrapped around.
The second reason is that we follow the accretion model, where near
periastron the secondary accretes mass and the secondary's wind is very
weak.

Our model reproduces to within a factor of two the results of
Hamaguchi et al. (2007), and to a good degree its qualitative behavior.
This is done without any parameters fitting. For example, if the
primary's wind is such that its mass loss rate is $2^{1/2}$ smaller,
and its wind velocity $2^{1/2}$ larger, than the values we use, then the
column density will be very close to the observed value. The primary's wind
momentum flux would stay the same, such that the X-ray properties of the
shocked secondary's wind would not change.
Our results clearly shows that from our preferred line of sight
($\omega=90^\circ$,$i=45^\circ$) the column density hardly changes during
most of the orbital cycle and can supply the required high column density,
in accord to observations. When the system approaches periastron passage
there is a fast increase of $N_{H,\rm{shock}}$ and $N_{Hi1}$, followed by a
decrease after periastron passage.

We note that the observations in phase 0.47 mentioned by Hamaguchi et
al. (2007), $N_H=17\times10^{22} \cm^{-2}$, occurred when the $2-10
\keV$ emission was $10-15 \%$ above its average value during that time
(Corcoran 2005). This could result from a higher density of the primary
wind. According to the model of Akashi et al. (2006) the dependance of
the X-ray emission on the primary mass loss is $L_x \sim \dot
M_1^{1/2}$. Namely, it is quite possible that the primary mass loss
rate, and hence $N_H$, was higher than the average value near apastron
by a factor of $\sim 1.25$, and this is the reason Hamaguchi et al. (2007)
did not find the column density to increase between phases
$0.47$ ($-0.53$)and 0.92 ($-0.08$).

Another effect not considered by us, and that introduces more variations,
both in the time variation and in the absorption by the conical shell
along different directions, is the corrugated structure of the shocked primary
wind that results from instabilities (Pittard \& Corcoran 2002; Pittard et al. 1998;
Okazaki et al. 2008a).

As mentioned earlier, from behind the secondary shock, it is not possible
to account for $N_H=17\times10^{22} \cm^{-2}$ near apastron.
It cannot come from the nebula, as the nebula can supply
$5\times10^{22}\cm^{-2}$ at most. We therefore conclude that the $N_H$
observations support a value of $\omega \simeq 90^\circ$.

\section{HIGH EXCITATION NARROW LINES}
\label{sec:narrow}

The narrow lines originate mainly in the Weigelt Blobs (WBs) located around the
main source (e.g. Damineli et al. 2008b; Nielsen et al. 2007b, Gull et
al. 2006, Smith et al. 2004). A representative high-excitation narrow line is
the [Ar III] $\lambda 7135$ line (Damineli et al. 2008b). As noted by Damineli
et al. (2008b), this line has many similarities to the radio light
curve at $3 \cm$ (Duncan \& white 2003), mainly by showing a continuous
variability along the cycle.
Although the main source of the line is the Weigelt blobs, contributions from
other regions is expected (Groh et al. 2007). This can be deduced from the presence of gas
between the WBs and to other directions at about the same distance (e.g.,
fig. 1 in Gull et al. 2006).

In an attempt to model this line we build a simple model based on the periastron-longitude
$\omega=90^\circ$ proposed by us.
We examine high-ionization lines, such that the ionization energy must
be supplied by the secondary and not by the primary.
We examine the rate of secondary ionizing photons that reach WB-D, as a function
of the orbital phase.
The location and angular dimensions of Weigelt blob D where measured from
Figure 1 of Gull et al. (2006), taking the distance to $\eta$ Car to be $2.3 \kpc$.
Blob D is located $0.25^{\prime \prime}$ from the central
source (Nielsen et al. 2007b; Gull et al. 2006). We assume that its location lies exactly
on the continuation of the semi-major axis of the elliptic orbit (See Fig.
\ref{fig:map}). Taking the inclination angle to be $45 ^\circ$, we find
blob D to be $r_D \simeq 850 \AU$ ($\sim 5$ light-days) from the primary.
With the secondary at the apex, Blob D forms a base of a cone with a full
opening angle of $\sim 30^\circ$ (Gull et al. 2006). We therefore examine the ionizing
radiation emitted within a cone with a half-opening angle of
$15^\circ$, and toward blob D. Blob B seems to be in the same direction as
blob D, and this model applies to emission from blob B as well.
However, there is a contribution to the emission-line intensity
from other regions, such as blob C. Therefore, the model will only show that the
periastron-longitude we propose can reproduce the general observed behavior. Detail
modeling and parameter fitting is postponed until after the next event.

The model is based on the one we built for the radio emission (Kashi \& Soker 2007a),
where the gas responsible for the radio free-free emission was assumed to be ionized
mainly by the secondary.
Whereas in the model for the radio emission the secondary ionizing radiation to all
directions was considered, presently we examine only the radiation toward the WBs,
and within a cone with a half opening angle of $15^\circ$.
For more detail on the model the reader should consult Kashi \& Soker (2007a).

In its way from the secondary to the blob, the secondary radiation suffers from absorbtion by the
post-shocked primary's wind gas in the conical shell, and by the undisturbed primary's wind.
The total absorbing rate per steradian is marked by $\dot{N}_{\rm abs}$.
We take the rate of secondary's ionizing photons entering the $15^\circ$ cone towards the blob,
and subtract from it the absorption rate.
Our results very close to periastron cannot be considered accurate for
the same arguments given in section \ref{sec:N_H}.
The primary's wind mass loss rate is $\dot M_1=3 \times 10^{-4} M_\odot \yr^{-1}$, and
its terminal speed $v_1=500 \km \s^{-1}$. From these we can calculate the absorption by the
undisturbed primary's wind.
The absorption by the post-shock gas in the conical shell depends strongly on its
density there, because the recombination rate of the gas depends on the density squared.
The density of the post-shock primary gas depends on the compression of the
post-shock gas by the ram pressure of the primary's wind.
The compression is determined by the balance between the ram pressure of the wind
and the internal pressure of the post-shock gas, including thermal and magnetic pressure.
Therefore, the compression strongly depends on the magnetic field strength and geometry in the
pre-shock primary's wind.
The magnetic field strength is characterized by the pre-shock magnetic to ram
pressure ratio $\eta_B$ (see Kashi \& Soker 2007a for more technical details).
This compression factor of the shocked wind introduces large uncertainties.

\begin{figure}[!t]
\resizebox{0.49\textwidth}{!}{\includegraphics{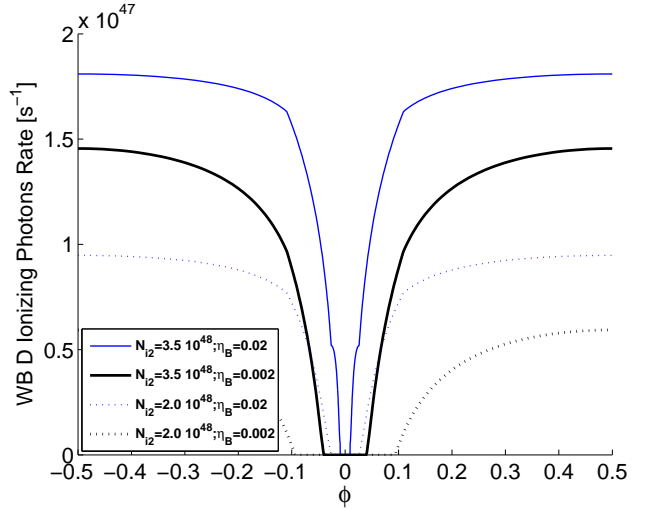}}
\caption{\footnotesize The rate of ionizing photons reaching Weiglet
Blob B and D ($\dot{\tilde{N}}_{iD}$), as function of the orbital
phase. Periastron occurs at $\phi=0$. Four cases are shown, according to
the value of the magnetic pressure to ram pressure ratio, $\eta_B$ in
the preshock primary's wind, and the rate of emission of ionizing
photons by the secondary, $\dot {N}_{i2}$ per steradian. }
\label{fig:WB_D_dependance}
\end{figure}

It is difficult to know the exact recombination rate close to the stagnation point ($\alpha \simeq
140^\circ-180^\circ$; see Kashi \& Soker 2007a). In that region, self-ionization may change
the recombination rate.
In order to partially compensate for this effect, we have made a cut-off to the conical shell's
increased recombination rate at $\alpha=140^\circ$. Namely, we assumed that the recombination rate per
steradian in the range $\alpha = 140^\circ-180^\circ$ is equal to the one calculated at
$\alpha = 140^\circ$.
This procedure causes a small change in behavior of the
calculated rate of ionizing photons reaching the blobs, like the ones seen in the upper line of
Figure \ref{fig:WB_D_dependance}, at $\phi=\pm 0.0264$ ($\theta=\pm125^\circ$).
The rate of ionizing photons at blob D is at minimum close to periastron because the dominant absorber
is the primary's undisturbed wind.
The column density between the secondary and the blob becomes larger as the secondary dives into the
primary`s wind acceleration zone close to periastron passage
(recall that the conical shell does not exist near periastron).

In the spirit of this paper, we minimize the parameter fitting, and consider only the most
influential processes.
For example, we do not consider the very likely possibility that the primary
wind is clumpy, and we neglect the instabilities in the wind interaction region
(Pittard \& Corcoran 2002; Pittard et al. 1998; Okazaki et al. 2008a).
In addition, we do not consider the ionization of the conical shell by the
radiation of the shocked winds, which might be significant near the
stagnation point. We stay with the same parameters that we used in our
model for the radio emission.
These are the pre-shock magnetic to ram pressures ratio, $\eta_B=0.001-0.1$,
and the rate of emission of ionizing photons per steradian by the secondary,
$\dot{N}_{i2}=2-3.5\times10^{48} \rm{s^{-1} sr^{-1}}$
(in a recent paper, Teodoro et al. (2008) obtained $\dot{N}_{i2}=2\times10^{48} \rm{s^{-1} sr^{-1}}$).
The rate of ionizing photons reaching blob D is given by
\begin{equation}
\dot{\tilde{N}}_{iD}=[\dot{N}_{i2}-\dot{N}_{\rm abs}]\Omega,
\label{eq:nid}
\end{equation}
where $\Omega= 0.214 \rm{sr}$ is the solid angle covered by the blob.
In Figure \ref{fig:WB_D_dependance} we show the dependence of
$\dot{\tilde{N}}_{iD}$ on $\dot{N}_{i2}$ and on $\eta_B$.

As expected, the ionizing flux reaching the blobs increases with the
increase in the secondary photon ionizing rate. Because the value of $\dot{\tilde{N}}_{iD}$
is the difference of two numbers, it might become sensitive to the value of
$\dot{N}_{i2}$ when the absorption rate $\dot{N}_{\rm abs}$ is non negligible.
This is clearly the case for our model.
As the strength of the pre-shock magnetic field, parameterized by $\eta_B$,
increases compression deceases. The lower density implies a lower photon absorption rate
$\dot{N}_{\rm abs}$, and the value of $\dot{\tilde{N}}_{iD}$ increases.

\begin{figure}[!t]
\resizebox{0.49\textwidth}{!}{\includegraphics{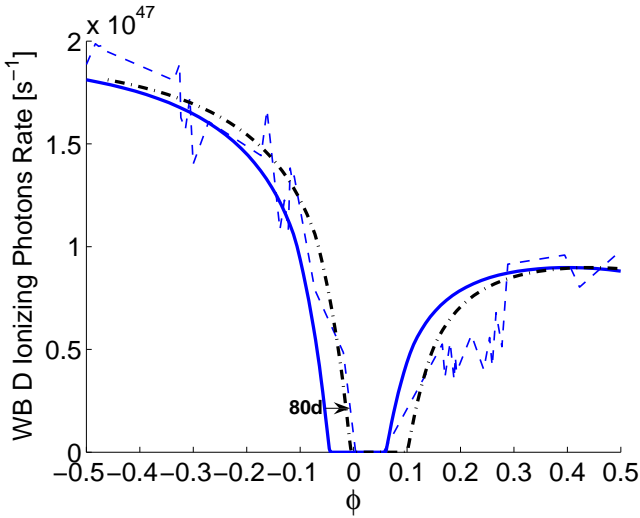}}
\caption{\footnotesize Our fit to the Weiglet Blobs B and D ionizing photons rate
(thick solid line) and the normalized intensity of the [Ar III] $\lambda 7135$
narrow dashed line (Damineli et al. 2008b; thin dashed line).
$\eta_{B}$ is given in equation (\ref{eq:etaB}), and
$\dot{N}_{i2}=3.5\times10^{48} \rm{s^{-1} sr^{-1}}$.
The fit is also drawn with $80 \days$ recombination time delay (thick dash-dotted line). }
\label{fig:WB}
\end{figure}

To reach a better fitting we now consider two effects that were
neglected in the simple calculation presented in Figure
\ref{fig:WB_D_dependance}. They are the likely variation of the
magnetic field with time, and the presence of gas around blobs D and B.
In Kashi \& Soker (2007a) we showed the magnetic field of the
primary's wind to be an important factor in the behavior of $\eta$ Car.
The problem with the magnetic fields is its stochastic nature, as is
well known from the sun. We have no knowledge of its behavior. But
knowing it is very important, we try to deduce its behavior in our
model. So basically, in our model the behavior of ionizing photons can
provide us some information about the magnetic field in the primary
wind.
>From the behavior of the sun we know that
the magnetic field can have semi-periodic,
as well as stochastic, variations. In addition, the magnetic field has spatial variations,
neglected by us. We consider the simplest variation possible, where the magnetic strength
of the primary's wind decreases from a maximum value during its active phase, taken to
be at phase $\phi=-0.5$ in the time period we consider, to a quiescent value.
The functional form we use is
\begin{equation}
\eta_B=\eta_{B-act}\exp(-t/\tau) + \eta_{B-q},
\label{eq:etaB}
\end{equation}
where $\eta_{B-q}$ is the quiescent value of $\eta_B$.
We have no prior knowledge of the values of these parameters, so we looked for
a simple fitting. We found that $\tau=1/6 \times \rm{cycle} \simeq 340 \days$,
$\eta_{B-q}=0.001$, and $\eta_{B-act}=0.004$ give good fitting.
The maximum value of $\eta_B$ occurs at $\phi=-0.5$ ($t=1012 \days$ and it is $\eta_B(max)=0.024$.
In Figure \ref{fig:WB} we plot the total number of ionizing photons
available to form narrow lines in blobs B and D, when the value of $\eta_B$ from
equation (\ref{eq:etaB}) is used.
For comparison we also plot the normalized intensity of the [Ar III]~$\lambda 7135$
narrow line, as given by Damineli et al. (2008a).

The matter residing in the vicinity of the blobs, as well as in other regions,
has two effects. Firstly, these regions are continued to be ionized even after photons
do not reach any more the WBs. Secondly, their lower density implies that
their recombination time is longer, and their decline can lag behind on a typical time scale
of weeks, e.g., 6 weeks for an electron density of $N_e=10^{-6} \cm^{-3}$, which
is $\sim 10$ times lower than the density in the blobs (Damineli 2008b).

These effects, i.e., recombination time of lower density regions, are important in the decline,
when the contribution of the WBs to the emission lines rapidly declines.
We find that a shift of the $\dot{\tilde{N}}_{iD}$ plot by 80 days to a later time
gives better fit to the observation just before the event.
This is presented by the dashed-dotted line in Figure \ref{fig:WB}.
Our explanation to the better fit with this 80 days displacement is that the contribution
to this line emission just before the event, when the emission is already weak, comes mainly
from regions with a recombination time of $\sim 80$~days.
It is evident that the fitting during most of the cycle, when emission is high,
does not change much, but the fitting during the decline phase is much better.

We can summarize this part by emphasizing the success of the model based on our proposed
periastron longitude.
By a simple model and fitting, the behavior of the high
excitation line emission can be reproduced if the absorption of the secondary
ionizing radiation by the shocked primary's wind in the conical shell is considered.
This can only be the case for a periastron longitude of $\omega \simeq 90^\circ$, namely,
the secondary is away from us during apastron and during most of the orbital motion
(see fig. \ref{fig:map}).
The main ingredients of the model are the same as that used by us in explaining the
radio emission from $\eta$ Car (Kashi \& Soker 2007a), and no new model had to be
invented here.
We only had to use the parts relevant to the WBs. The consideration of the
variation of the magnetic field, for example, was also used in the radio model.

\section{DISCUSSION AND SUMMARY}
\label{sec:diss}

\subsection{Main results}

Our goal was to learn about the orientation of the semimajor axis of the
$\eta$ Car binary system from several different observations.
We found that an orientation where the hotter secondary star is closer to us for
a short time at periastron, i.e., a periastron longitude of $\omega=90^\circ$,
fits these observations.
During most of the time the primary is closer to as, as depicted in Figure \ref{fig:map}.

In section \ref{sec:doppler} we attributed the high excitation He~I lines to the secondary
wind, and by that could reasonably fit the variation of the Doppler shift with the
orbital motion (for more detail see KS07).
The Doppler shift of the low excitation Fe~II~$\lambda6455$ line could be fitted by
attributing its origin to the primary stellar wind.
The orbital motion explanation for the Doppler shift accounts also for the
absence of the Doppler shift variation toward the polar directions.
The doppler shift of the Paschen lines could be accounted for if they are assumed to
be formed in the shocked primary's wind near the stagnation point (see Fig. \ref{fig:map}).
We also note that the variation in the Doppler shift of the X-ray lines might
also be caused by the orbital motion (Behar et al. 2007)

Trying to explain the lines' behavior, we used two sets of primary and secondary masses
which we consider possible (see KS08) and obtain nice fits without over-playing with other parameters.
We emphasize that claims for exaggerated values for those masses are no problem
for the qualitative model we present. Using lower masses we were still able to fit all the
lines in the paper by slightly adjusting other parameters (eccentricity, inclination,
magnetic-compression, cone opening angle, mass-loss rates, etc.) well within their acceptable range.

In section \ref{sec:N_H} we examined the hydrogen column density toward the hard
X-ray emitting gas, $N_H(>5 \kev)$.
The column density at several times along the orbit is given by Hamaguchi et al. (2007).
The column density is sensitive to the mass loss rate and velocity of the primary's wind,
and to the nature of the wind interaction process, e.g., where exactly the gas emitting the
hard X-ray resides.
We cannot reproduce the exact variation of $N_H$ with orbital phase, but could reproduce
the approximate value at each phase (Fig. \ref{fig:NH}).

In an opposite binary orientation, where the secondary
is toward us during most of the time (near apastron), the expected value of $N_H$
near apastron is much smaller than the observed value. It is also expected to rise
toward periastron by a much larger factor than what is observed.
Over all, although our fit is not perfect and requires further work, it
has less severe problems than what a model based on an opposite orbital
orientation would have.

We did not deal with the value of $N_H(1 \keV)$ toward the gas emitting the soft X-ray
($\sim 1 \keV$), as we are not sure where this gas is located.
However, during most of the time $N_H(1 \keV) \simeq 0.3 N_H(>5 \kev)$.
This suggests that if it was not for the dense primary's wind around the secondary,
the hot X-ray emitting gas would have had lower $N_H$ than the observed value.
It is not easy to account for the high $N_H$ value in a model where during most of
the time we observed the shocked secondary's wind through the tenuous secondary
wind ($\omega \sim 270^\circ$).

In section \ref{sec:narrow} we calculated the hard ionizing radiation that is emitted
by the secondary and reaches the Weigelt blobs (WBs).
We found that when the absorption by the undisturbed and shocked primary's wind is considered,
the qualitative behavior of the high excitation [Ar III]~$\lambda 7135$ line that
is assumed to be emitted mainly by the WBs (Damineli et al. 2008b) can be reproduced.

The absorption by the shocked primary's wind$-$ in the conical shell$-$depends on the
compression of the post-shock gas (Fig. \ref{fig:WB_D_dependance}).
In our model (Kashi \& Soker 2007a) the compression is constrained
by the post-shock magnetic pressure.
The post-shock magnetic pressure is determined by the ratio of the pre-shock
magnetic pressure to the wind's ram pressure $\eta_B$.
To fit the observed behavior (Fig. \ref {fig:WB}) we had to assume that the
magnetic field in the primary's wind evolves according to
equation (\ref{eq:etaB}).
In general, it is expected that the magnetic field in stellar wind will change over time
scales of years to tens of years, as it is very well known for our sun.

To summarize, using the model that has been proposed in our study of the radio emission
from $\eta$ Car (Soker \& Kashi 2007a), we reproduced the basic properties
of the ionizing radiation that are required to form the high excitation narrow lines.

\subsection{Further considerations}

We list four observational results that our proposed
$\omega=90^\circ$ periastron longitude can account for.

\subsubsection{The evolution of the radio emission}

The evolution of the radio emission also supports the orientation
proposed here. The radio image contains some bright knots, with fainter
emission in areas between and around these knots (Duncan \& White
2003). In particular we note a bright radio knot to the same direction
relative to the center as the WBs are. If the orientation was opposite
to what we claim here, then the ionizing radiation of the secondary
toward this bright region would be constant until $\la 3$~months before
periastron. The reason is that the radiation from the secondary to the
knot would propagate through the tenuous secondary's wind. Namely, in
the model where the secondary is toward us near apastron the bright
radio region would stay more or less at the same maximum brightness
until phase $\sim -0.05$, while region to the sides would decline
slowly, as the secondary dives into the denser part of the primary's
wind. However, a careful examination of the radio movie (White et al.
2005) shows that the radio knot fades after apastron passage toward the
1998 and 2003 minima, as the rest of the nebula does. This is expected
in the periastron longitude $\omega=90^\circ$ proposed by us, when the
knot is being irradiated through the primary's wind, which continuously
becomes denser as the system approaches periastron. This fading of the
radio emission as a result of the secondary `diving' into the dense
primary's wind has been discussed before by Duncan \& White (2003) and
Abraham \& Falceta-Gon\c{c}alves (2007).

\subsubsection{The purple haze}
Smith et al. (2004) presented the ultraviolet images of the Homunculus at 6 epochs around the
2003 minimum.
They found that just before periastron the bright UV region extended to the south-east, along the
symmetry axis of the Homunculus.
Smith et al. (2004) assumed that the UV emitting-reflecting regions are in the equatorial
plane, and from that they deduced that the semi-major axis is perpendicular to our line of sight,
namely, a periastron longitude of $\omega=0$.

Abraham \& Falceta-Gon\c{c}alves (2007) also attributed
the excess UV radiation to the secondary star.
Close to periastron this radiation, according to {Abraham \& Falceta-Gon\c{c}alves (2007),
is confined by dust absorption to the interior region of the colliding-winds cone;
the cone points eastward before, and westward after, the event.

We suggest a different interpretation of that behavior.
We note that the south-east lobe is toward us, and attribute the UV bright region in
the south-east direction to the polar direction, rather than to the equatorial region.
For example, a transient polar outflow (Behar et al. 2007) that opens a cone for
the ionization and UV illumination might be the cause of this illumination,
rather than the orbital orientation at this phase.

We also note the following interesting result: The region of the Weigelt blobs
became UV-bright for a short time about a month before periastron passage
(Smith et al. 2004).
This is at the same time as the peak in the narrow He~I line intensity;
the He~I peak is seen a month before the 2003.5 minimum (event 11), but not in the
previous two minima (Damineli et al. 2008b).
According to the orientation proposed in our model, the secondary is toward this side (closer to us) near
that time, and it is possible that for a short time the ionizing radiation from the
secondary toward the blobs actually increased, e.g., due to clumpy primary's wind or
instabilities an opening in the dense conical shell was formed.

\subsubsection{The He~I $\lambda10830$ line}

The He~I $\lambda10830$ high excitation line has a P Cygni profile with
absorption changing from $-640 \km \s^{-1}$ to $-450 \km \s^{-1}$
(Damineli et al. 2008b). Just before periastron a wide wing in
absorption appears, reaching $\sim - 1000 \km \s^{-1}$ a month before
periastron, and $\sim - 1400 \km \s^{-1}$ at periastron. This shows
that there is a flow of a high velocity gas toward us at periastron.
The primary star is not expected to blow such a fast wind toward us,
and if it does, why it cannot be seen in other phases as well? We
attribute this behavior to the shocked secondary's wind or polar
outflow. In our proposed orientation the secondary is closer to us at
periastron. The shocked secondary's wind will flow toward us near
periastron passage. This is why the wide wing of the He~I
$\lambda10830$ line is seen only close to periastron passage.
In any case, as suggested by the recent study of Teodoro et al. (2008), the
interpretation of the He~I $\lambda10830$ requires careful attention,
and will be the subject of a forthcoming paper.

\subsubsection{The location of the Weigelt blobs}

Chesneau et al. (2005) observed the sub-arcsec butterfly-shaped dusty environment surrounding $\eta$ Car
with the VLTI, using the mid-IR instrument MIDI, and the adaptive Optics system NACO.
As shown in their figure 7, the lower density `SE filament' (named so by Smith et al. 2003a)
is apparently aligned in the same direction as the NW Weigelt blobs complex.

The exitance and the large mass of the
Weigelt blobs might be related to their location close to the periastron of the
system orbit, where the secondary's wind is strongly disturbed by the
primary's dense wind.
The counterpart (the SE direction) of this high density region is a very low density region,
but has some filaments inside it.
In the NE direction has even a lower density.
The higher density in the NW might be the consequence of the eccentric orbital motion of the
secondary.
We predict that 3D numerical simulations using our proposed orientation
will reproduced this density asymmetry.

\subsection{Unclosed issues}

The proposed orbital orientation and our study might have some weak points:

(1)\emph{Coincidence.} The coincidence that the secondary's
wind velocity where the lines are formed, $v_{\rm zone} \simeq -430 \km
\s^{-1}$, is practically the same as the primary wind terminal
velocity. Although this is a somewhat weak point in our model, our
answer to that coincident (KS07) is that the same can be said about the
model where the He~I lines originate in the primary's wind: How come
the changes in the velocity of the regions where the lines are formed
in the primary's wind exactly mimic the secondary velocity around the
center of mass? Another coincident is that the best orientation is
exactly for the secondary to be toward us at periastron, and not even
several degrees from that direction. However, we found (KS07) that
several degrees deviation from $\omega=90^\circ$ is possible. Some
other weak points regarding the orbital motion interpretation for the
Doppler shifts are listed by Davidson (1997). However, like Damineli
(1997) he attributed the He~I lines to the primary, while we attribute
them to the secondary.

(2) \emph{Magnetic field.} The absorption of ionizing radiation depends
strongly on the magnetic field in the primary's wind. Our need for
magnetic field adds a parameter to the model. As it stands now it is a
weak point. However, if magnetic fields are detected in the primary
wind, then this becomes a strong point. In particular, the magnetic
field becomes strong in the shocked primary's wind. We encourage a
search for magnetic fields in the shocked primary region, e.g., by
looking for its influence on some lines. The strength of the magnetic
field near the stagnation point is expected to be $B_{\rm shock} \sim
100 (r/1 \AU)^{-1}$~G, if we assume that the post-shock magnetic
pressure is about equal to the wind ram pressure. This field can be
detected, e.g., in the Paschen lines very close to periastron passage.

We note the following. Our need for magnetic field comes from the following consideration.
If the post-shock primary gas has no magnetic field, it is compressed to a very high density
in the conical shell, such that it absorbs too much of the ionizing radiation.
Instabilities in the conical shell can reduce this absorption even without magnetic fields,
because most of the gas might be concentrated in dense clumps, and the ionizing
radiation escapes between this clumps. the study of this process
requires 3D numerical simulations.

(3) \emph{Helium lines from the secondary's wind.} The secondary
luminosity is only $\sim 20 \%$ of the total luminosity. This might
cause some problems in attributing the He~I lines to its wind. However,
we note that most of the radiation from $\eta$ Car is in the IR anyhow.
A detail study of the line formation in the secondary is required,
similar to the one Hillier et al. (2001, 2006) conducted for the
primary star. For the time being we note that many stars similar to the
secondary are known to have P-Cygni He~I lines with velocity much
smaller than their terminal velocity (see discussion in KS07).

\acknowledgements

We thank Otmar Stahl, Arnout van Genderen, and Diego Falceta-Gon\c{c}alves,
and an anonymous referee for enlightening comments.
This research was supported by the Asher Space Research Institute in the Technion,
by P. and E. Nathan research fund, and by the Israel Science Foundation (grant No. 89/08).

\end{document}